\documentclass{svjour3}                     
\smartqed  
\usepackage{graphicx}
\usepackage{mathptmx}      
\usepackage{natbib}
\journalname{SSRv}
\begin{document}

\title{Plasma Turbulence in the Local Bubble\thanks{This research was supported at the University 
of Iowa by grant ATM03-54782 from the National Science Foundation}
}

\titlerunning{Local Bubble Turbulence}        

\author{Steven R. Spangler}
\institute{Department of Physics and Astronomy, University of Iowa, 
           Iowa City, Iowa 52242, USA\\
              Tel.: 001-319-335-1948\\
              Fax: 001-319-335-1753\\
              \email{steven-spangler@uiowa.edu}  }

\date{Received: date / Accepted: date}

\maketitle

\begin{abstract}
Turbulence in the Local Bubble could play an important role in the thermodynamics of the gas that is there. This turbulence could also determine the transport of cosmic rays and perhaps heat flow through this phase of the interstellar medium.  The best astronomical technique for measuring turbulence in astrophysical plasmas is radio scintillation.  Scintillation measurements yield information on the intensity and spectral characteristics of plasma turbulence between the source of the radio waves and the observer.  Measurements of the level of scattering to the nearby pulsar B0950+08 by Philips and Clegg in 1992 showed a markedly lower value for the line-of-sight averaged turbulent intensity parameter $<C_N^2>$ than is observed for other pulsars, qualitatively consistent with radio wave propagation through a highly rarefied plasma.  In this paper, we discuss the observational progress that has been made since that time. The main development has been improved measurements of pulsar parallaxes with the Very Long Baseline Array.  This provides better knowledge of the media along the lines of sight.  At present, there are four pulsars (B0950+08, B1133+16, J0437-4715, and B0809+74) whose lines of sight seem to lie mainly within the local bubble. The mean densities and line of sight components of the interstellar magnetic field along these lines of sight are smaller than nominal values for pulsars, but not by as large a factor as might be expected.  Three of the four pulsars also have measurements of interstellar scintillation.  The value of the parameter $<C_N^2>$ is smaller than normal for two of them, but is completely nominal for the third.  This inconclusive status of affairs could be improved by measurements and analysis of ``arcs'' in ``secondary spectra'' of pulsars, which contain information on the location and intensity of localized screens of turbulence along the lines of sight. Similar data could be obtained from observations of highly compact extragalactic radio sources which show the ``intraday variability'' phenomenon.    
\end{abstract}

\section{Introduction}
The Local Bubble is a medium of low density and high temperature, and thus is a plasma medium.  The interstellar magnetic field doubtlessly threads through this region, making magnetized plasma phenomena important.  We would like to diagnose the usual plasma parameters in this medium as well as the properties of turbulence, if it exists.  These properties would include the ``wave mode'' which most naturally describes the excitations, the energy density of turbulence, and its spatial power spectrum. 

Such knowledge could be important in better understanding the dynamics of the Local Bubble, and the role it plays in the general galactic interstellar medium.  Turbulence has the potential of dissipating on fast timescales; if the energy density of the turbulence is comparable to other energy densities, turbulent dissipation can play an important role in the heating of this medium, as it seems to do in the solar corona as well as other parts of the interstellar medium. 

Turbulence in the Local Bubble could also play an important role in the galactic transport of cosmic rays.  At times in the past it has been thought that ``tunnels'' like the Local Bubble could fill a large fraction of the volume of the interstellar medium.  If the mean free path of cosmic rays in these bubbles or tunnels is very long, longer than in other parts of the interstellar medium, these bubbles might play a major role in the transport of cosmic rays throughout the Galaxy.  In a collisionless plasma, turbulent fluctuations will determine parameters such as the pitch angle and spatial diffusion coefficients, and effectively the mean free path of particles.   
\section{Radio Astronomy and Diagnostics of Interstellar Plasmas}
The refractive index of radio waves in a plasma depends on the density of the plasma and, to a much lesser but nonetheless observable degree, the magnetic field.  As such, radioastronomical propagation measurements can yield information on plasma characteristics such as the plasma density and magnetic field. Path-integrated measurements of the plasma density can be obtained from measurement of the dispersion measure (DM). Observationally, the dispersion measure describes the degree to which radio waves of different frequencies travel at different speeds. Physically, it is the path integral of the electron density along the sight.   It is only measurable for pulsars.  The rotation measure (RM) is also a path integral along the line of sight.  Observationally, it is a measure of how the polarization position angle changes with wavelength, and it can be measured for pulsars or extragalactic radio sources. The integrand in the RM is the product of the electron density and the line-of-sight component of the magnetic field.    

Of equal interest, radio waves propagating through a turbulent medium with random density fluctuations undergo a number of modifications in amplitude and phase which are collectively described as interstellar scintillations (ISS).  This topic was last extensively reviewed by \cite{Rickett90}.  The major ISS phenomena which have been observed and analysed are shown in cartoon form in Figure 1.  
\begin{figure}
\includegraphics[width=0.60\textwidth]{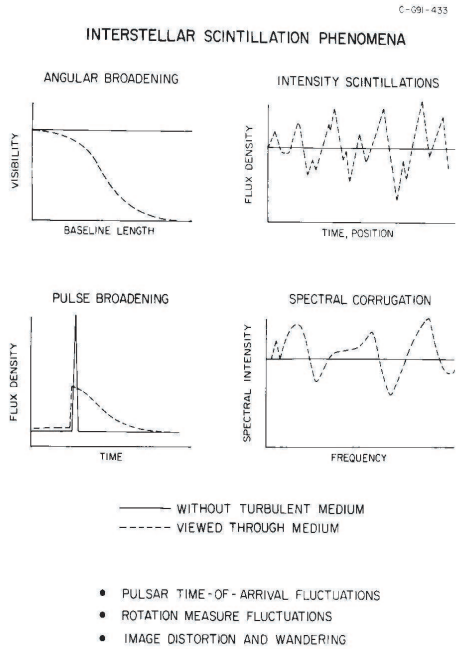}
\caption{The variety of ISS phenomena.  The solid lines indicate what is measured if the source is viewed without an intervening turbulent screen.  The dashed lines indicate what is observed in the presence of turbulence. For the representation of angular broadening we show the measurement of a radio interferometer.  The magnitude of the interferometer visibility (the properly normalized correlation coefficient of the wave electric field at two antennas) is plotted versus the baseline length between the interferometer elements. The term ``spectral corrugation'' refers to a stochastic variation in the flux density as a function of radio frequency, induced by wave passage through a turbulent medium.  Figure taken from \cite{Spangler07}.  }
\end{figure}
The different scintillation observations illustrated in Figure 1 yield different information about the properties of interstellar turbulence. 

In general, ISS observations yield information on the path integral of density turbulence properties, with different observational techniques featuring different kernels in the path integral.  However, as will be discussed below, ISS observations sometimes contain additional information such as the distribution of turbulent plasma along the line of sight. 

Some of the phenomena illustrated in Figure 1 can be observed for either pulsars or extragalactic radio sources.  Others are observable only for pulsars.  In this paper, I concentrate on pulsars because the lines of sight are confined within the Galaxy, and in the most interesting case would lie completely within the Local Bubble. 

It is typical in interstellar scintillation observations to model the density spatial power spectrum by 
\begin{equation}
P_{\delta n}(q) = \frac{C_N^2}{\left[ q_z^2 + \Lambda^2(q_x^2 + q_y^2 )\right]^{\alpha/2}}
\end{equation}
where $q_x,q_y$, and $q_z$ are three Cartesian components of the spatial wavenumber.  The index $\Lambda$ describes the anisotropy of the irregularities. The power spectrum is rotationally symmetric about the $z$ axis.  In many analyses, an isotropic spectrum has been assumed in which $\Lambda=1$, and such will be assumed for most of this paper. However, ISS observations have presented compelling evidence that the interstellar density spectrum is anisotropic; a recent summary of the evidence for this anisotropy is given in \cite{Rickett07}.  Anisotropy of the turbulence is expected theoretically on the basis of recent theories of magnetohydrodynamic (MHD) turbulence \citep{Spangler99,Chandran02}.

The parameter $C_N^2$ can be related to more intuitive properties of plasma turbulence via the following formula, which relates $C_N^2$ to the mean electron density $\bar{n_e}$, the ``modulation index'' of the turbulence $\epsilon \equiv \frac{\delta n}{\bar{n_e}}$, and its outer scale $l_0$ \citep{Spangler98}
\begin{equation}
C_N^2 = f(\alpha)\left[ \frac{\epsilon^2}{(1+\epsilon^2)l_0^{2/3}}\right]\bar{n_e}^2
\end{equation}
In equation (2) $f(\alpha)$ is a dimensionless number which depends on $\alpha$; for the Kolmogorov value $\alpha=11/3$, $f(\alpha)=0.181$.
Analyses of interstellar scintillation data attempt to retrieve $C_N^2$, $\alpha$, and $\Lambda$. Most ISS observations yield an estimate of $C_N^2$ which is averaged over the line of sight.  This observational quantity is referred to as $<C_N^2>$. The discussion in this paper will concentrate on the value of $<C_N^2>$ which can be retrieved from published observations of lines of sight which have passed through the Local Bubble.  I am particularly interested in comparing $<C_N^2>$ for lines of sight which are largely within the Local Bubble to those which pass through the normal ISM in the vicinity of the Sun.  
\section{Why the Hot Phase of the Interstellar Medium is of Interest}
What can selected pulsar observations tell us about plasma turbulence in the local bubble? What makes this part of the interstellar medium (ISM) of particular interest?  One response is that it has plasma parameters which may differ in important ways from the plasma in the more extensively diagnosed Diffuse Ionized Gas \citep[DIG;][]{Cox87}.  One of the more important such parameters is the plasma beta, which may be defined as 
\begin{equation}
\beta \equiv \frac{c_s^2}{V_A^2}
\end{equation} 
where $c_s \equiv \sqrt \frac{5/3 k_B (T_e + T_i)}{m_p}$ is the ion acoustic wave speed, and $V_A$ is the Alfv\'{e}n speed.  In the expression for the ion acoustic speed, $T_e$ and $T_i$ are the electron and ion temperatures, and $m_p$ is the proton mass.  Choosing nominal Local Bubble parameters of $n_e \simeq 5 \times 10^{-3}$ cm$^{-3}$,  $T_e = T_i \simeq 10^6$ K, and a standard ISM magnetic field strength of $B=4 \mu G$, we have  $\beta = 2.2$.  This is larger than the corresponding value for the DIG of $\beta \simeq 0.4$.  It should be noted that the magnetic field strength in the Local Bubble may be lower than we have assumed here.  \cite{Opher06} estimate $|B| \simeq 2 \mu G$ immediately outside the heliosphere.   Such a value for the magnetic field in the Local Bubble would increase our estimate to  $\beta \simeq 9$.  Finally, it is possible that the magnetic field strength is lower still in the truly low density Local Bubble, as opposed to the immediate vicinity of the Sun.  This would result in an even higher estimate for $\beta$ in the Local Bubble. 

The plasma $\beta$ plays a very important role in plasma wave and turbulence dynamics.  Damping mechanisms such as Landau damping can depend strongly on the plasma $\beta$ \citep[e.g.][]{Barnes66}.  Nonlinear processes such as parametric decay instabilities also depend on the beta, being allowed for certain values and forbidden for others.  As a result, the dynamics of waves and turbulence could be quite different in hot phase plasmas like the Local Bubble relative to the DIG and colder phases.  The difference in the properties and dynamics of waves and turbulence in a high $\beta$ plasma could result in  similar differences in the interaction between turbulence and charged particles in the Local Bubble.    
\section{Probing the Local Bubble: the Case of Pulsar B0950+08}
A difficulty in interpreting radio scintillation observations is that a measurement has contributions from gas all along the line of sight.  Since most parts of the interstellar medium are much denser than the Local Bubble (particularly the DIG, which appears to be correlated with scintillations), the Local Bubble will typically make a small and inseparable contribution to the measured signal.  It has been clear for a long time that the key to remote sensing of the Local Bubble is to find a radio source which scintillates (in one of the observables shown in Figure 1) and whose line of sight is completely or mainly within the local bubble.  

It was clear 20 years ago that the pulsar B0950+08 was the best, and possibly only suitable object for this type of measurement. Based on a then-current parallax distance of 127 parsecs, the line of sight should have been entirely within the local bubble. 

Several years later, \cite{Phillips92}  published observations of B0950+08 showing interstellar scintillation, and deduced a value of $<C_N^2> = 2.82 \times 10^{-5}$ m$^{-20/3}$. This was the lowest value for $<C_N^2>$ that had been reported at that time, and was qualitatively consistent with expectations for a line of sight that was mainly within a low density bubble. By way of context for this measurement, the extensive survey of \cite{Cordes85} gives a typical value of  $<C_N^2>$ in the diffuse, or ``Type A'' scattering medium of about $<C_N^2>=3.16 \times 10^{-4}$ m$^{-20/3}$.  In the remainder of this paper, I will typically quote the base-10 logarithm of $<C_N^2>$, which for B0950+08 and the Type A medium are -4.55 and -3.5, respectively. 

Nonetheless, the value reported by \cite{Phillips92}, while about a factor of 10 lower than the average value of $<C_N^2>$ for the interstellar medium, was larger than might have been expected.  The basis of this statement is the formula for $C_N^2$ in equation (2).  The electron density in the local bubble is at least a factor of 5 less than that in the DIG (line-of-sight average of 
$5 \times 10^{-3}$ cm$^{-3}$ rather than  $2-3 \times 10^{-2}$ cm$^{-3}$ for the DIG), and the outer scale $l_0$ is probably larger than that in the DIG, so $<C_N^2>$ should arguably be even smaller than that reported by \cite{Phillips92}.  The fact that it is not might indicate that the Local Bubble, despite being of a lower electron density than the general interstellar medium, is more turbulently ``agitated''.  A more mundane possibility is that the measurements of B0950+08 are ``contaminated'' by denser plasma along the line of sight (see Section 8).     
\section{Relevant Recent Progress in Pulsar Astronomy}
It has now been sixteen years since the paper by \cite{Phillips92}, and we should have a clearer idea of the properties of turbulence in the Local Bubble.  There are several developments that are particularly relevant in this respect.  
\begin{enumerate}
\item The most important development has been the completion of the Very Long Baseline Array (VLBA) of the National Radio Astronomy Observatory\footnote{The Very Long Baseline Array is an instrument of The National Radio Astronomy. NRAO is a facility of the National Science Foundation, operated under cooperative agreement with Associated Universities Inc.}, and the full achievement of its astrometric capability \citep{Brisken00,Brisken02}.  This has permitted precise parallaxes to be measured for many pulsars.  The distance to a pulsar obviously allows us to determine what regions of the interstellar medium lie along the line of sight. 
\item The number of pulsars known has increased from 558 \citep{Taylor93} at the time of \cite{Phillips92} to 1775 at the writing of this paper.  A catalog of the properties of all known pulsars is maintained by the pulsar group at the Australia Telescope National Facility (ATNF) under Richard Manchester in a published form 
\citep{Manchester05} as well as an updated internet site (http://www.atnf.csiro.au/research/pulsar/psrcat/).  This catalog is an important research aid in its own right, and provided data for this paper. 
\item \cite{Lallement03} have published a three-dimensional map of the Local Bubble, using NaD absorption data (a diagnostic of the neutral phase) for 1005 lines of sight to relatively nearby stars.  This study has provided an accurate representation of the structure of the bubble. 
\item During this period there has been steady progress in our understanding of pulsars in the galactic environment, including the information they contain about the interstellar medium.  A prime example of a relevant recent discovery which has particular promise is discussed in Section 9 below.    
\end{enumerate}
\section{A New Set of Pulsars to Probe the Local Bubble}
We used the ATNF catalog to select a set of pulsars which can be used to probe the properties of the plasma in the Local Bubble.  The information extracted from the catalog included, most importantly, the parallaxes measured by \cite{Brisken00,Brisken02}.  The criteria used in selecting these pulsars were as follows.  
\begin{enumerate}
\item The distance to the pulsar had to be less than 500 parsecs.  
\item The part of the sky in which the pulsar is observed was required to be in the direction of the Local Bubble.  
\end{enumerate}
The purpose of these selection criteria was to select pulsars whose lines of sight are, to the greatest possible extent, within the Local Bubble.  The desired goal was to find a pulsar whose line of sight lies entirely within the Local Bubble, so that its radio propagation measurements would diagnose the plasma in that region.  

The pulsars so selected are given in Table 1.  Column 1 gives the pulsar name and Column 2 gives its distance. The parameters in the remaining columns are described in the next section.  In addition to distance and direction information, measurements of dispersion measure (DM) and Faraday rotation measure (RM) were taken from the ATNF catalog.  Data on ISS parameters were taken from elsewhere in the literature. The final row gives parameters for the Type A scattering medium, to which the properties of the Local Bubble lines of sight are compared.  
\begin{table}
\caption{Pulsar Probes of the Local Bubble}
\begin{tabular}{lllllll}
\hline\noalign{\smallskip}
Pulsar & Distance (pc) & $\bar{n_e}$ cm$^{-3}$ & $\bar{B}_{\parallel} \mu$G & $\log <C_N^2>$ & Corr. &  $\log <C_N^2>_A$\\
\noalign{\smallskip}\hline\noalign{\smallskip}
B0950+08 & 262 & 0.011 & 0.57 & -4.55 & A+B & -5.42 \\
B1133+16 & 350 & 0.014 & 1.00 & -3.32 & B & -3.82\\
J0437-4715 & 170 & 0.016 & 0.70 & \ldots & \ldots & \ldots  \\
B0809+74 & 433 & 0.014 & 2.37 & -5.16 & none & -5.16 \\
Type A medium & \ldots & 0.025 & 2.0 & -3.50 & A+B & -4.38 \\
\noalign{\smallskip}\hline
\end{tabular}
\end{table}
\section{Results on Plasma and Plasma Turbulence in the Local Bubble}
Using the new, VLBA-provided data on pulsar distances in Table 1, we can use published measurements of propagation and scattering effects to obtain measurements of path-averaged plasma properties such as the plasma density (from the pulsar dispersion measure), line-of-sight component of the magnetic field (from Faraday rotation measure), and $<C_N^2>$ (from ISS phenomena such as spectral corrugation or pulse broadening). 

The dispersion measure (DM) is defined as 
\begin{equation}
DM \equiv \int_{LOS}n_e ds = \bar{n_e}L
\end{equation}
If $n_e$ is expressed in cm$^{-3}$ and $ds$ in parsecs, DM has the customary, non-cgs units of pc-cm$^{-3}$. The variable $L$ is the effective path length (parsecs) through the medium.  The Faraday rotation measure (RM) can be expressed as \citep{Minter96}
\begin{equation}
RM \equiv 0.81 \int_{LOS}n_e \vec{B} \cdot \vec{ds}
\end{equation}
with $n_e$ and $ds$ again in cm$^{-3}$ and parsecs, and $|B|$ in microGauss.  The odd coefficient 0.81 is the transformed version of the conventional set of fundamental constants which appears in Faraday rotation formulas, when this set of units is chosen.  The units of $RM$ are rad/m$^2$.  

An estimate of the path-averaged, line-of-sight component of the interstellar magnetic field is 
\begin{equation}
\bar{B_{\parallel}} = \frac{\int_{LOS}n_e \vec{B} \cdot \vec{ds}}{\int_{LOS}n_e ds} = 1.24 \frac{RM}{DM} 
\end{equation}
In equation (6) $\bar{B_{\parallel}}$ is in microGauss. 
Formulas (4) and (6) are used to obtain $\bar{n_e}$ and $\bar{B_{\parallel}}$, which are given in columns 3 and 4 of Table 1. Measurements of scintillation parameters (such as the angular width of a broadened image) are proportional to a quantity called the scattering measure ($SM$) which is defined in a manner similar to dispersion measure and rotation measure, 
\begin{equation}
SM \equiv \int_{LOS} C_N^2 ds =  <C_N^2> L
\end{equation} 
In the simplest case,  $<C_N^2>$ is obtained by dividing the measured SM by the path length through the medium.  The process is usually more complicated in the case of pulsars, which are embedded in the ISM.  

The values for $<C_N^2>$ were taken from three different sources for the three pulsars with ISS measurements.  These are \cite{Phillips92} for B0950+08, \cite{Bhat98} for B1133+16, and \cite{Rickett00} for B0809+74.  The published values taken from these papers are given in column 5 of Table 1.  

These values cannot be directly compared with each other, or with the Type A background value of $<C_N^2>$, as I now discuss. The measurements of B0950+08 by \cite{Phillips92} were made in the same way, and analysed with the same formulas as \cite{Cordes85}; the factor of 10 difference in $<C_N^2>$  between B0950+08 and the general Type A background appears to be a sound result.  

\cite{Bhat98} made repeated ISS observations of all pulsars in their sample, with each pulsar being observed from 10  - 90 times.  They found that the fundamental ISS observable, the decorrelation bandwidth ($\Delta \nu_d$), varied markedly from one observing session to another.   The reasons for this variability are not entirely clear, but appear to be due to distortion of diffractive ISS processes by refractive ones.  To minimize this variability, \cite{Bhat98} averaged the two dimensional autocorrelation functions from the dynamic spectra to obtain a Global Average Correlation Function (GACF) prior to fitting for the decorrelation bandwidth. With the resulting value of $\Delta \nu_d$, they used the same formula as \cite{Cordes85} to obtain a value of $<C_N^2>$.  

As discussed by \cite{Bhat98}, and as is clearly visible in Table 2 of their paper, there is a systematic bias in the values of $<C_N^2>$ relative to those of \cite{Cordes85}, in the sense that the values of $<C_N^2>$ from the GACF correlation functions are lower.  This bias is presumably due to the aforementioned, poorly-understood refractive bias.  We calculated the difference in $\log <C_N^2>$ between \cite{Bhat98} and \cite{Cordes85} for the 14 pulsars in common, and obtained a mean value of 0.37, with a standard deviation of 0.35.  If  $<C_N^2>$ data from \cite{Phillips92} and \cite{Cordes85} are to compared with \cite{Bhat98}, this correction for refractive bias (hereafter indicated by the unimaginative name ``Correction A'') must be applied to the former data sets.  

The value of  $<C_N^2>$ for B0809+74 was taken from \cite{Rickett00}.  Scintillation parameters were obtained from measurements of intensity scintillations at a frequency of 933 MHz, rather than the autocorrelation function of a dynamic spectrum. \cite{Rickett00} obtain a smaller value for the scattering measure, and thus  $<C_N^2>$, from these data than had previous observations.  \cite{Rickett00} argue that previous investigations, including \cite{Phillips92}, \cite{Cordes85}, and \cite{Bhat98} had used an inappropriate value for the multiplicative constant in a formula which converted from $\Delta \nu_d$ to $<C_N^2>$ (equation (6) of \cite{Cordes85}).  To correct $<C_N^2>$ for this effect, \cite{Rickett00} suggest multiplying the published values of $<C_N^2>$ by a factor of 0.315, or adding -0.50 to $\log <C_N^2>$.  I refer to this as ``Correction B''.  

These corrections are reported in columns 6 and 7 of Table 1.  Column 6 indicates which (if any) correction should be applied to the published value of $<C_N^2>$ to obtain a consistent value.  Column 7 gives the corrected value of $<C_N^2>$. The adjusted values are noted by $<C_N^2>_A$.  The values in this column are, one hopes, directly comparable.  

The following conclusions may be drawn from Table 1.  
\begin{enumerate}
\item The crucial pulsar B0950+08 is now known to be twice as far as indicated by earlier parallax measurements.  This result comes from the improved parallax measurements of \cite{Brisken00}.  This means that much of the line of sight lies beyond the probable back wall of the Local Bubble, and renders the implications of the B0950+08 data for Local Bubble plasma physics much less certain. 
\item The tremendous increase in the number of known pulsars over the last 15 years did not assist the present endeavor.  Most of the new pulsars were found in the direction of the galactic center, and had evaded earlier detection because heavy interstellar scattering broadened the pulses sufficiently to make the pulsars undetectable.  The Local Bubble, on the other hand, lies in the direction of the galactic anticenter, where scattering is much less pronounced and earlier surveys had found the pulsars which can be detected.  
\item Turning to the plasma parameters in Table 1, the mean electron densities $\bar{n_e}$ are lower than the standard value for long lines of sight through the interstellar medium ($\bar{n_e} \simeq 0.025$), but they are substantially larger than expected for the hot gas in the Local Bubble, which would be $n_e \simeq 0.005$. This probably results from the lines of sight being ``contaminated'' by denser plasma outside the Local Bubble. 
\item Intriguingly, the inferred values for the line of sight component of the magnetic field are only slight less, or completely consistent with, lines of sight through the general interstellar medium. 
\item The scattering for two of the pulsars (B0950+08 and B0809+74) is less than on typical lines of sight through the ISM.    However, $<C_N^2>$ is not depressed by as much as one would expect on the basis of the lower density in the bubble.  Furthermore, one of the three pulsars (B1133+16) has a completely ``nominal'' scattering measure and associated value of $<C_N^2>$.  In considering the scattering measurements, we are again confronted with the question of how much of the measured propagation effects are due to plasmas outside the Local Bubble, and how much is due to plasma within the Local Bubble.   
\end{enumerate}
\section{Estimates of Other Media Along Pulsar Lines of Sight} 
From the summarizing points of the previous section, we conclude that either the Local Bubble has higher density, turbulent intensity, etc, than expected, or the measurements are being contaminated by other media along the lines of sight.  In this section we briefly discuss what we know of these other media, and estimate the magnitude of the contamination.  

The first medium to consider is the denser ISM behind the Local Bubble.  The lengths of the lines of sight given in Table 1 are larger than estimates of chord lengths in the Local Bubble. To estimate the degree of contamination by the ISM beyond the Local Bubble, we used the meridian slices through the Local Bubble and its vicinity published in  \cite{Lallement03}.  The slices shown in Figures 6-8 of \cite{Lallement03} were used to plot the length and orientation of each pulsar's line of sight.  In no case did the line of sight remain entirely within the Local Bubble.  It appears to be the case that no known pulsar is sufficiently close, and in the right direction, to be entirely within the Local Bubble.  Typically, about half of the line of sight appears to be in the denser (presumably typical) ISM behind the far wall of the Local Bubble.  
These rough estimates appear consistent with the result from Table 1 that the line-of-sight averaged densities are about half of what is typically observed for pulsars on longer pathlengths. They are also consistent with the normal, or nearly normal, values of $\bar{B_{\parallel}}$ for the pulsars in Table 1. 

The values of $<C_N^2>_A$ given in Table 1 for B0950+08 and B0809+74 appear to be lower, relative to the adopted mean for the Type A medium, by more than a factor of 0.50 (0.30 in the logarithm).  The portion of the line of sight from the far wall to the pulsar is at distances above the galactic plane of order 100 parsecs and greater for both pulsars.  The intensity of interstellar turbulence at these altitudes may be less than in the galactic plane.  Furthermore, the portion of the line of sight in the denser ISM behind the wall is between the pulsar and the midpoint of the line of sight.  The properties of intensity scintillations are dependent on the location of the turbulence along the line of sight; turbulent plasma close to the pulsar has less of an effect than plasma at the midpoint.  

The other medium known to be on the lines of sight to these pulsars consists of clouds relatively close to the Sun and embedded in the Local Bubble, such as the LIC cloud and G cloud. These have been discussed recently by \cite{Frisch07} and \cite{Redfield08}.  To estimate the contributions of these clouds to DM, RM, and SM for the pulsars, we used the mean cloud properties given in Section 4.2 of \cite{Redfield08}.  We assume that these clouds have an electron density $n_e = 0.12$ cm$^{-3}$ \citep{Redfield08b}, radius $R_c = 1.5$ parsecs, and that they fill space out to a distance of $L_c = 15$ parsecs from the Sun with a filling factor $0.055 \leq f \leq 0.19$. 
Rotation measures were calculated with the same assumed parameters, and an adopted value of $B_{\parallel} = 4 \mu G$.  Since the magnitude of the interstellar magnetic field in the DIG is estimated to be $ 4 \mu G$, the calculated RMs will be upper limits.   

The turbulence parameter $C_N^2$ was calculated via equation (2), using the same estimate for $n_e$ as used for DM and RM.  The outer scale to the turbulence $l_0$ in the local clouds is totally unknown, but it is not unreasonable to assume that it is a fraction of the radius of the clouds, so we adopt $l_0 \simeq R_c/3 \simeq 0.5$ pc. Equation (2) shows that the value of $C_N^2$ is not strongly dependent on the assumed value of $l_0$.  The modulation index $\epsilon$ is also an ``imponderable''.  In the absence of other information, we adopt values measured for the density turbulence in the solar wind.  \cite{Spangler02} and \cite{Spangler04} obtained estimates for this quantity in the slow solar wind, utilizing radio remote sensing observations for the solar wind in the heliocentric distance range 16-26 $R_{\odot}$, and spacecraft measurements at 1 au.  Based on these results, we adopt an estimate of $0.050 \leq \epsilon \leq 0.15$.  

With these assumed parameters, our estimates for the contribution of the local clouds to the pulsar propagation parameters are as follows.  Calculated dispersion measures are in the range $0.10 \leq DM \leq 0.34$ pc-cm$^{-3}$, depending on the assumed value of the filling factor.   The observed dispersion measures for the pulsars in Table 1 range from 2.6 to 6.1  pc-cm$^{-3}$ \citep{Manchester05}.  The local clouds might make a small contribution to the observed dispersion measure, but most of DM must come from the local bubble or the ISM beyond the Local Bubble.  

Estimated local cloud values for RM are in the range 0.32 - 1.11 rad/m$^2$, again dependent on the assumed filling factor.  The observed RMs have absolute values between 1.35 and 11.7 rad/m$^2$, with three of four pulsars having $|RM| < 4$ rad/m$^2$ \citep{Manchester05}.  These RM estimates should be considered upper limits; the true values will be reduced by the unknown but generally nonzero angle between the line of sight and the magnetic field in the clouds.  Nonetheless, it is possible that the local clouds could make a significant contribution to the rotation measures for these pulsars.  

Our estimates of the turbulence parameter $C_N^2$ in the clouds are $-4.0 \leq \log C_N^2 \leq -3.0$.  These values are mostly larger than those given in column 7 of Table 1, although a direct comparison is misleading.  The values of  $<C_N^2>_A$ given in Table 1 are averages over paths of several hundred parsecs. Since the effective path lengths through the local clouds are very small compared to the typical pulsar distance, the effective value of  $<C_N^2>_A$ due only to these clouds would be much smaller than the local value of  $C_N^2$. To roughly estimate this correction, we multiplied the above, calculated values of  $C_N^2$ by a factor of $fL_c/D$. This is equivalent to taking the SM due to the clouds and dividing by the distance to the pulsar.  These estimates of the effective cloud values of  $<C_N^2>_A$ show a very wide range, reflecting uncertainty in both the cloud filling factor as well as the modulation index $\epsilon$.  The maximum possible value is comparable to the observed  $<C_N^2>_A$ for both B0950+08 and B0809+74. The observed scattering to B1133+16 is larger than can be plausibly accounted for by the local clouds. 

This exercise indicates that the observations of ISS for two of the pulsars in Table 1 (B0950+08 and B0809+74) are not incompatible with a significant role for turbulence in the set of local clouds described by \cite{Frisch07} and \cite{Redfield08}. It is worth noting that such a role is also strongly indicated by the interpretation of the intraday variability of three compact quasars by \cite{Linsky08}.  

This section may be summarized as follows. The four pulsars in Table 1 furnish our best chance of diagnosing the plasma and plasma turbulence in the Local Bubble by radio propagation effects. However, the lines of sight to these pulsars pass not only through the Local Bubble, but also the system of small clouds close to the Sun and the general ISM beyond the far wall of the Local Bubble. The estimates presented in this section indicate that the contributions of these two media to DM, RM, and SM, separately as well as together, are at least plausibly consistent with the measured values.  While further investigation is certainly warranted, it is difficult to reject the null hypothesis that all of the radio propagation effects occur in media outside the Local Bubble.

\section{New Diagnostic Measurements for Local Interstellar Plasmas}
The previous section remains inconclusive because we cannot be sure how much of an observed propagation effect is due to plasma within the Local Bubble, and how much is due to (arguably less interesting) plasma outside the bubble. 
In the last few years, new observational advances have been made which hold the prospect of progress.  A particularly promising new observational development is the discovery of ``arcs'' in pulsar ``secondary spectra''.  Secondary spectra are defined in the next paragraph. This intriguing arc phenomenon was discovered by Daniel Stinebring and his students at Oberlin College. Recent expositions and physical discussions of the phenomenon are given by \cite{Walker04} and \cite{Cordes06}.  

For many years, pulsar astronomers have studied pulsar dynamic spectra, which are two dimensional plots of the pulsar intensity as a function of frequency and time, $I(\nu,t)$.  For at least 20 years, it has been realized that these plots often showed variations which are not entirely random, but instead consisted of patterns which suggested constructive and destructive interference between a few rays. 

Something remarkable was noted when Stinebring and colleagues examined the ``secondary spectrum'', which consists of Fourier transforming the dynamic spectrum.  The secondary spectrum may be considered a plot of the intensity as a function of the conjugate variables to frequency and time, $f_{\nu}$ and $f_t$ respectively.  What was found for many pulsars is most of energy is concentrated on parabolic arcs described by the simple relation 
\begin{equation}
f_{\nu} = af_t^2
\end{equation}
In general, pulsars have more than one such parabolic arc in the secondary spectrum.  Clear examples are shown, for example, in Walker et al (2004) and Cordes et al (2006).  These papers also demonstrate that such behavior is easily understood as interference between a non-scattered central ray from the pulsar to an observer, and a scattered secondary ray.  Each pair of rays produces its own parabolic arc.  

For these arcs to be present, it is necessary for there to be relative motion between the medium containing the turbulent irregularities, the pulsar, and the observer.  It also appears to be necessary, rather unexpectedly, for the turbulent density fluctuations to be concentrated in relatively narrow sheets or screens along the line of sight.  For many years the ``thin screen approximation'' has been employed in the theory of wave propagation in a random medium, in which the turbulent medium is idealized as being concentrated in a small fraction of the line of sight.  It now seems to be the case that this approximation is physically accurate for the interstellar medium.  

The coefficient of the parabolic arc contains information on the location of the screen responsible for the pulsar arcs.  The coefficient $a$ in equation (2) may be shown to be  \citep{Cordes06}
\begin{equation}
a = \frac{Ds(1-s)\lambda^2}{2cV_{\perp}^2}
\end{equation}
where $D$ is the distance to the pulsar, $s$ is the fractional distance of the screen along the line of sight (i.e. $sD$ is the distance from the pulsar to the screen), $c$ is the speed of light, and $V_{\perp}$ is the speed of the observer relative to the scintillation pattern, in a direction perpendicular to the direction of the pulsar. The wavelength of observation is $\lambda$.

The discovery of pulsar arcs opens the possibility of determining what fraction, if any, of the scattering of the pulsars in Table 1 is due to turbulence in the Local Bubble.  One of the pulsars, B1133+16, has prominent arcs in its secondary spectrum.  These data and their analysis are discussed by Trang and Rickett (2007).  The curvature $a$ of the primary arc gives a screen distance of 140 parsecs from Earth, which may place it within the Local Bubble. The charts of \cite{Lallement03} are not entirely clear as to whether a point on the line of sight to B1133+16 140 parsecs from the Sun is within the Local Bubble, or in the denser surrounding medium.  If it can be determined that this point is within the Local Bubble, the ISS observations of \cite{Trang07} are of great interest to the topic of this paper. 

Observations of ISS for B0950+08 and B0809+74 have been published, but the data do not seem to have had secondary spectra calculated, and analyses of the sort of interest here have apparently not been made.  The archived data could be reanalysed with this in mind, or new observations could be made.  
In view of the suggestive results for B1133+16, it would be highly desirable to search for pulsar arcs for all the pulsars in Table 1. These data could indicate what part of the scattering is due to plasmas in the Local Bubble, and what part is due to more remote plasmas. 

Another technique with considerable promise is that of intraday variability of highly compact extragalactic radio sources \citep{Rickett07}.  A disadvantage of extragalactic radio sources with respect to pulsars is that the lines of sight pass all the way out of the Galaxy, so there is no prospect of having a line of sight begin and end within the Local Bubble.  A compensating advantage of extragalactic sources is their high surface density in the direction of the galactic anticenter and the Local Bubble, where very few pulsars are to be found.  Furthermore, a model-dependent estimate of the distance to the scattering plasma can be made from these intensity scintillation measurements.  Intraday variability observations of three radio quasars were used by \cite{Linsky08} to indicate that the turbulence responsible for radio scattering comes from the ionized edges of a number of local clouds within the Local Bubble. A unique aspect of the \cite{Linsky08} study was the ability to identify the three dimensional velocities of the various clouds with the observed ``pattern speed'' of the scintillations with respect to the radio telescope. 

It is worth noting that the successful modeling of the quasar intraday variability by \cite{Linsky08} indicates that the density fluctuations responsible for this component of ISS are confined to sheets or screens in the local clouds or boundaries between local clouds.  This conclusion is similar to that drawn for the pulsar arcs discussed above, and may reveal an important and general property of astrophysical turbulence. In view of the success of the intraday variability technique, further observations should be carried out.  Once again, it would be interesting to determine how much of the scattering measure can be attributed to these clouds, and how much might be contributed by other plasmas in the Local Bubble and beyond.  
\section{Summary and Conclusions}
\begin{enumerate}
\item There are at least four pulsars which are potential probes of the plasma in the Local Bubble.  These pulsars are listed in Table 1. 
\item Even these pulsars are probably too far away to have their lines of sight entirely ``fit'' within the Local Bubble. Unfortunately, it seems to be the case that there are no radio pulsars which are close enough to have the entire line of sight within the bubble. 
\item Two of the three pulsars with ISS observations have lower values of scattering measure, and thus the scattering parameter $<C_N^2>$, than is typical for pulsars without special lines of sight. 
\item A significant, or even dominant portion of the observed dispersion measure, rotation measure, and scattering measure values to these pulsars may arise in the denser ISM beyond the far wall of the Local Bubble, or in the local clouds in the vicinity of the Sun. The local clouds may contribute a significant portion of the observed ISS.  
\item The one pulsar which has had its scintillation data analysed for the presence of ``pulsar arcs'' (B1133+16) has a prominent arc that might be caused by a turbulent screen within the Local Bubble. 
\item If most of the scattering of B1133+16 is due to plasma within the Local Bubble, it would present a challenge to our understanding of interstellar turbulence in two ways.  First, the intensity of the turbulence, parameterized by the coefficient $<C_N^2>$ is unexpectedly high.  Second, it is hard to understand how or why plasma turbulence in the hot, high $\beta$ plasma of the Local Bubble would be concentrated in a thin sheet within the bubble.  
\item The remaining three pulsars in Table 1 should be examined for arcs in their secondary spectra, either by reanalysing archival data, or by making new observations. 
\item Additional, targeted observations of highly compact extragalactic sources for intraday variability could provide similar information on the location of turbulent plasma which might be located within the Local Bubble.   
\end{enumerate}

\begin{acknowledgements}
This work was supported by the National Science Foundation through grant ATM03-54782 to the 
University of Iowa. 
\end{acknowledgements}

\end{document}